# A Conversation with Robert V. Hogg

**Ronald Herman Randles**

*Abstract.* Robert Vincent Hogg was born on November 8, 1924 in Hannibal, Missouri. He earned a Ph.D. in statistics at the University of Iowa in 1950, where his advisor was Allen Craig. Following graduation, he joined the mathematics faculty at the University of Iowa. He was the founding Chair when the Department of Statistics was created at Iowa in 1965 and he served in that capacity for 19 years. At Iowa he also served as Chair of the Quality Management and Productivity Program and the Hanson Chair of Manufacturing Productivity. He became Professor Emeritus in 2001 after 51 years on the Iowa faculty. He is a Fellow of the Institute of Mathematical Statistics and the American Statistical Association plus an Elected Member of the International Statistical Institute. He was President of the American Statistical Association (1988) and chaired two of its winter conferences (1992, 1994). He received the ASA's Founder's Award (1991) and the Gottfried Noether Award (2001) for contributions to nonparametric statistics. His publications through 1996 are described in *Communications in Statistics—Theory and Methods* (1996), 2467–2481.

This interview was conducted on April 14, 2004 at the Department of Statistics, University of Florida, Gainesville, Florida, and revised in the summer of 2006.

## THE EARLY YEARS

**Randles:** Bob, what can you tell us about the years before you entered the field of statistics?

- **Hogg:** I went through the first grade in Hannibal. It was during the Depression. Dad had a chance to go with Consolidated Coal as a traveling coal salesman where he would go around to retail stores and sell coal to various dealers. We had to move to Rockford, Illinois, which in a sense probably was a good thing, because the school system at that time was better in Rockford than in Hannibal. It was about sixth grade that I realized that I was better in mathematics than the other kids. I had a sixth grade teacher named Miss Davis, and I definitely remember her giving some story problems that I could solve and the others kids couldn't. At that time I caught on to the spirit of the thing and by the time I went into junior high I wanted to get all A's. Then in high school I had a good math teacher, Katherine Slade, and she spurred me on. Of course, World War II came along right at that time; I was a senior on December 7, 1941. So we knew that we were going into military service. Actually, I did finish a year in college before I entered the military. I was in the Navy V12 college program. Dad never gave me a lot of advice, except he had been in World War I in the infantry, received a silver star and purple heart, so he was a good soldier. But when I was enlisting, I could sign up for either the Navy's V12 or the Army's ASTP program. Dad told me to choose the Navy because I wouldn't have to be in the trenches. That was a good piece of advice, because the Navy recognized that the war was going to go on for a long time and they wanted to keep these young men in college to become officers later. The Army's ASTP program folded after a semester or two and some of my friends who were in that program were taken off to fight in Europe. While I didn't get a college degree under the V12 program, I'd received enough that after the war when I went back to school, I only had about a year left.

*Ronald Herman Randles is Professor, Department of Statistics, University of Florida, Gainesville, Florida 32611-8545, USA (e-mail: rrandles@stat.ufl.edu).*

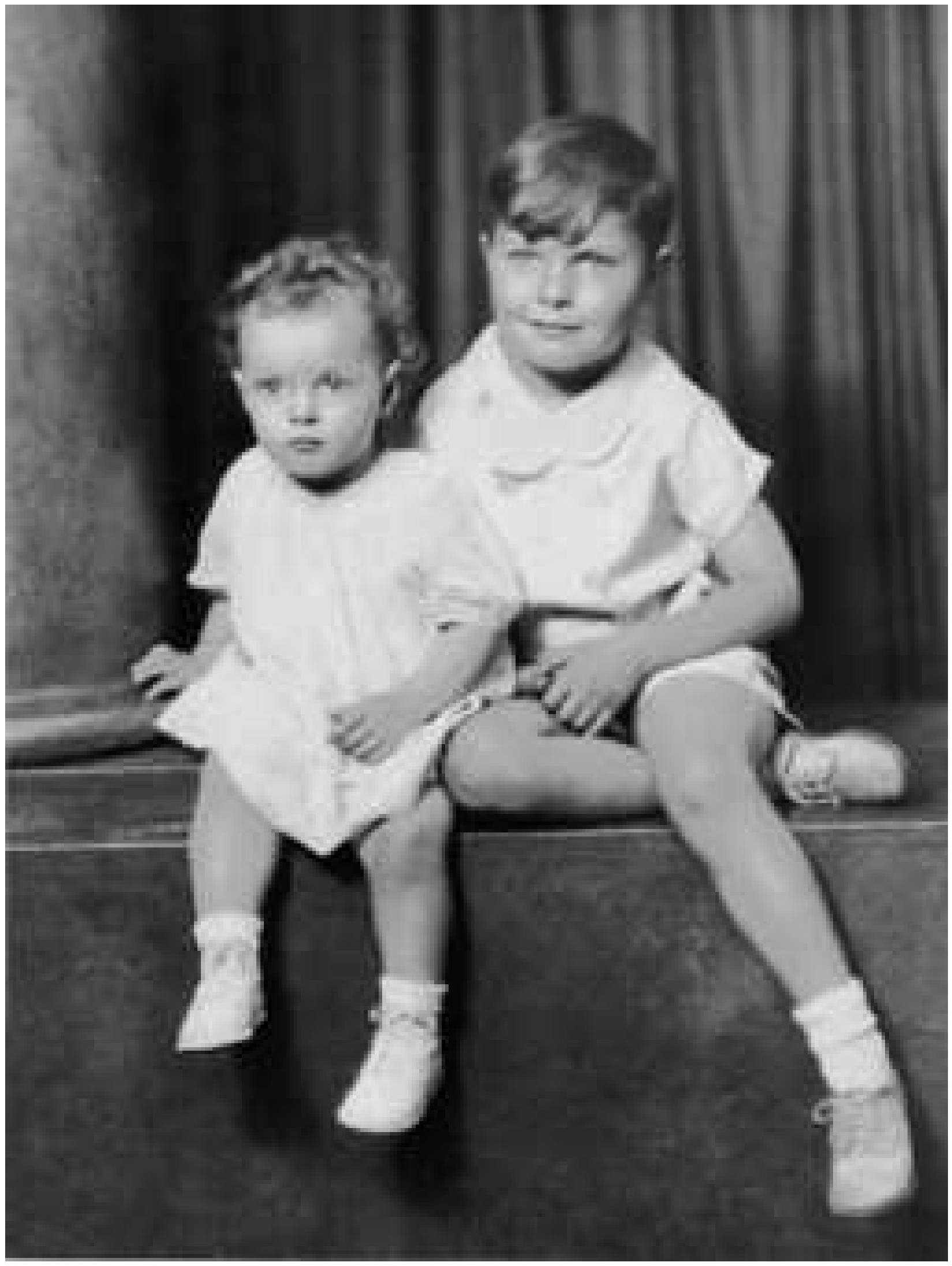

Fig. 1. *Bob Hogg, age 4 1/2, with sister, Ginger.*

**Randles:** How long were you on active duty?

**Hogg:** Oh, from '43 to '46, essentially three years. When I was discharged in July, 1946, we were living in Rockford, Illinois at the time, so I went to the University of Illinois and graduated in 1947, in math. I took probability in the spring out of Uspensky, taught by an advanced graduate student named Evans Monroe who was a Ph.D. student of Joe Doob's. He got me interested in probability. I remember there were two other students in that class and they too had been veterans. We walked out of that probability class one day, it was probably March or April, and we wondered, "What are we going to do with a math major?" It was in the basement of the math building at the University of Illinois and we saw this sign, "Do you want to be an actuary? Come to the State University of Iowa." By the way, it was the State University of Iowa at that time. The reason we dropped "The State," Iowa State College became a university in the early '60s and people got confused. They were Iowa State University and we were the State University of Iowa, so we quietly dropped "the State" at that time. After the three of us saw that sign, we applied for assistantships. We were essentially straight A students and so all three of us went over to Iowa. The other two graduated from Iowa and became actuar-ies. But during the second semester I had fallen in love with statistics taught by Allen T. Craig. I asked him, "If I stayed on and studied a little more math, could I work with you?" He said "sure." So I took much more mathematics. We had a small math department, only 12 faculty members. There weren't too many courses to take. But I took courses from a professor named Edward Chittenden who was pretty well known in math and a man named Malcolm Smiley in algebra. I got reasonably good training, but I was working with Craig on a statistics problem. I only had six hours of the theory of statistics, which was maybe a little harder than Hogg and Craig. As I think back, Ron, that was a great period. As a Ph.D. student that last year, I used to go in to see Allen a lot, more than any Ph.D. student ever saw me. I can't imagine him spending all those hours with me, but we really hit it off. When I was going to get my degree in 1950 the academic jobs had dropped way off. Right around '46, '47, '48 when the veterans came back Iowa went up to 12,000 students, but in 1950, when I was getting out, it had dropped to 7000, which was the enrollment pre-World War II at Iowa. But every school had the same experience. So jobs just didn't exist. You know I probably could have gotten a job at a much smaller school, but Allen suggested to the Chairman of the Department of Mathematics that they hire me on and they did.

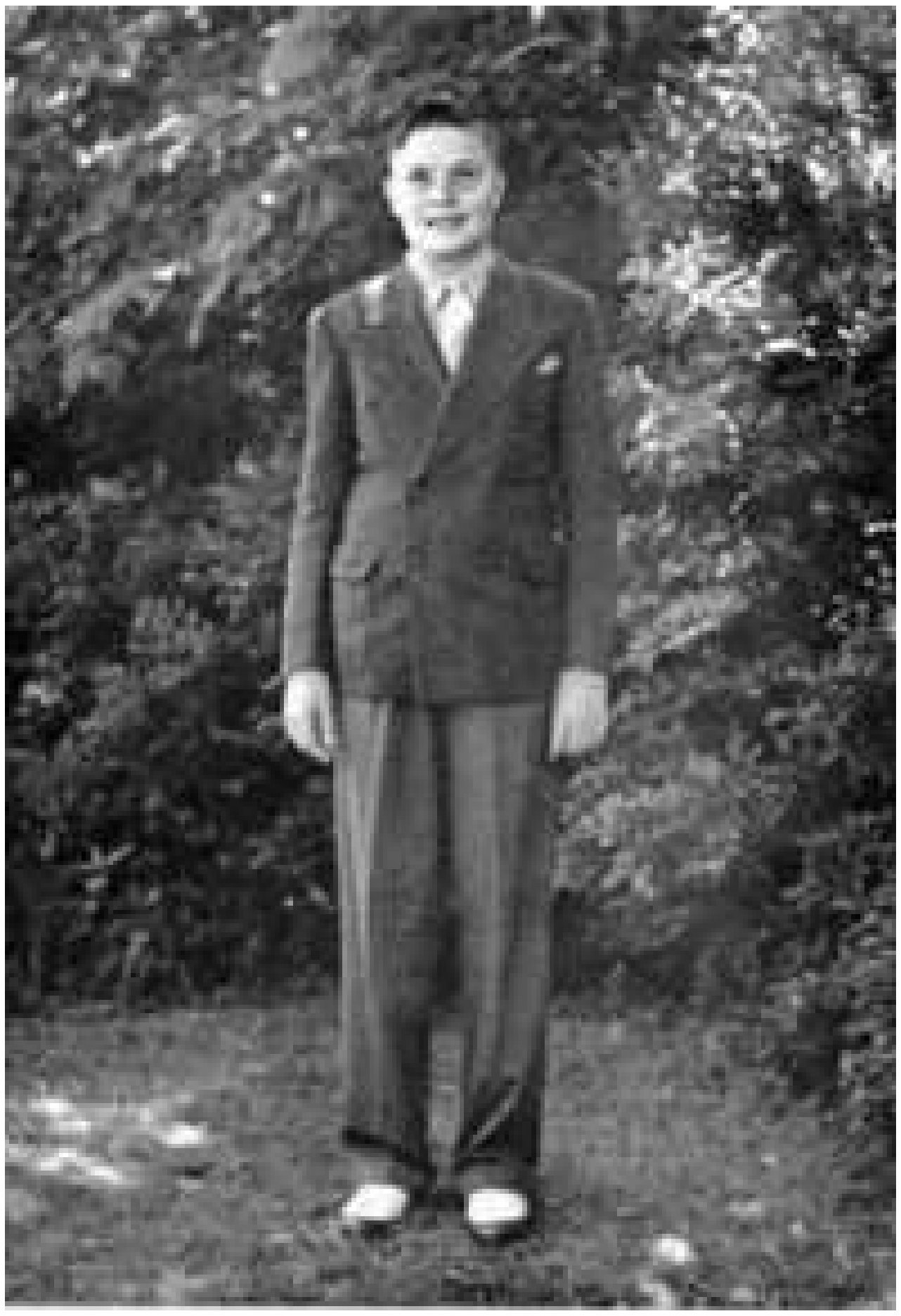

Fig. 2. *Bob Hogg age* 14.

**Randles:** So you began teaching statistics at the University of Iowa in 1950.

**Hogg:** Yes, I taught two calculus classes which were five-hour engineering calculus classes and one three-hour stat course. We participated in a seminar too. So I was up to 14 semester hours of teaching that first semester. As a matter of fact, for the first year or two I taught that much each semester. Then this algebraist, Malcolm Smiley, suggested that we "ought to give the younger guys a break." Right after that I was down to two courses. Sometimes I might teach a calc course and a stat course and maybe seven or eight hours or something like that, but essentially it was down to more or less the six-hour load. Thank goodness Malcolm did that. You know it's interesting that as I look back, I was single at the time, and I was spending lots of time down at the building. I would go down there almost every night. Probably I had more time then teaching 13 or 14 hours than I had later on teaching six or seven. At least it seemed that way.

What we did in seminars was wonderful. We would study books. I wanted to study more statistics. The first book we picked out was a great selection. It was Cramér [4]. In those seminars, we had three faculty members: Craig, an actuary Byron Cosby and me, plus two or three students. We would study a certain section each week. We all studied: it wasn't as if somebody went to the board and lectured on it and the rest of us hadn't done the work. But I tell you that the faculty would study it harder than the students because we didn't want to look bad. As I recall, we actually worked on Cramér for two years. We studied books during the fifties like Fraser, Lehmann, Anderson and Scheffé's books [1, 5, 17, 18]. Fraser was one of the first advanced nonparametrics books and Lehmann, while on testing of hypotheses, had some nonparametrics in there, but from both the books there was one thing I learned that was very important. If the sample arose from a continuous distribution, the order statistics were complete and sufficient statistics for the distribution. That would serve me well later on.

## THE HOGG AND CRAIG TEXT

**Randles:** The Hogg and Craig textbook [11] is considered a classic in mathematical statistics. Many statisticians received their first theory course from it through the years. What prompted the development of this important textbook?

**Hogg:** Craig and I were very interested in sufficient statistics. We taught what later appeared in Hogg and Craig basically through the '50s. Initially we were thinking of doing a monograph on sufficient statistics. We didn't like the way any book presented distribution theory. So in the late '50s after we had been teaching the material for several years, we decided to write it up as a book and it took about a year to complete it. I look back on that first edition, which was a little thin edition and probably more of a preliminary edition than a first edition in all honesty, but we had presented sufficient statistics by page 100. One of our first things after we wrote the book, Berkeley adopted it and we thought that we had it made. We did a second edition in '65 and probably that was much more polished. But sufficient statistics got pushed back a little bit. We were very innovative with the treatment of sufficient statistics. I've talked to people who were studying statistics in the '60s, in the '70s and a lot of them had Hogg and Craig. It is really amazing.

**Randles:** The use and application of sufficiency was certainly one of the major innovations of the Hogg and Craig text. The text was also prominently known for its treatment of the change-of-variable method of deriving the distribution of a function of several random variables.

**Hogg:** The change-of-variable and functions of random variables material is good. We were thinking of a monograph on sufficient statistics, but we didn't like the way anybody did change of variables and functions of random variables. In the first chapter we covered some discrete and continuous distributions, in the second chapter some conditional probability, the third chapter had some special distributions, but it was standard material. But the fourth chapter, we did the change of variable and that was a great chapter. We were doing that primarily to get into the sufficient statistic material. While I say the innovative part was the use of the sufficient statistic, our transformation of variables, in all honesty, was very carefully written. Chapter 4 was a key chapter.

## ALLEN CRAIG

**Randles:** Bob, tell us more about Allen Craig, with whom you worked so closely and who was not only your teacher and mentor but a good friend through the years. I know you named your oldest son Allen, after Allen Craig, because of the close relationship between the two of you.

**Hogg:** Allen was from Florida. He went to the University of Florida (where this interview was taking place). He was a joint math and classics major. He never married. Back then, Florida wasn't as good a university as it is today. Allen said he was lucky in the fact that he had a few teachers who had to come to Florida for health reasons and some of those professors were very good. Actually how did Allen decide to go to Iowa? I am really not sure, except Henry Reitz was a very well-known actuary and statistician. Allen went up to Iowa in the late '20s. Just to show how Henry Reitz attracted students, Sam Wilks went to Iowa at the same time and he was from Texas. You have to figure that somebody was drawing them. Iowa also had this professor Ed Chittenden, who was a pretty well-known analyst, although not a real good teacher. He attracted some people, like Dean Montgomery, who was his most famous student, and another good one, Neal McCoy. Chittenden and Reitz attracted some excellent students. Reitz really started the actuarial program. Later on he founded the Institute of Mathematical Statistics in 1935. Harry Carver had started *The Annals of Mathematical Statistics* in 1930. But it was really Reitz who founded the IMS and Craig then became its first Secretary-Treasurer in 1935. Then Sam Wilks became the first Editor of *The Annals of Mathematical Statistics* after Carver in 1938. Sam did a magnificent job for about 10–12 years. He took it from a so-so journal to a world-class statistics journal. Those three, Henry Reitz, Allen Craig (who was one of Sam's Associate Editors when he took over) and Sam Wilks really had a lot to do with the beginnings of the Institute of Mathematical Statistics. So Allen was very active in this.

When Allen received his Ph.D. from Iowa in '31, Sam Wilks did too. Reitz decided to keep Allen on as a teacher. Allen was a better lecturer than Wilks, but Wilks was a better statistician than Allen. Wilks went on to Columbia and then to England for a year or so and then came back to Princeton. Allen Craig volunteered for service in World War II. Allen would have been in his late 30s or early 40s. He was eventually assigned to a destroyer. In 1943 Reitz fell into bad health. Sam, being Editor of the *Annals*, wanted to dedicate the '43 *Annals* to Reitz, because everybody knew Reitz was going to die. By the way, it is kind of a neat thing that Sam did because they dedicated the 1943 *Annals* to him before he actually died. He tried to get some of Reitz's students to write up articles. Allen contributed two articles. One of them was on order statistics and it wasn't too important. But the other was on $AB = 0$, the Craig theorem for independence of quadratic forms. Allen was on a destroyer at the time and all his notes were back in Iowa. Going one way, if $AB = 0$, it's child's play to show the independence as the joint moment-generating function or the characteristic function factors. Going the other way is tricky. This is not typical Craig at all, but Allen waved his hands and tried to go backward. But it was published [2]. Hotelling noted the error in going that way. He tried it, but Hotelling messed up the proof also. Craig finally published the proof that he had after the war. It was published in '47 [3]. There he was considering the independence of two bilinear forms. So he was in a multivariate situation, but it was the same proof that he had. We give the proof in Hogg and Craig, but we use one by H. O. Lancaster.

Allen was a great southern gentleman. He was very polite. When Carolyn and I got married, he was part of the wedding party. He was the godparent of each of our children. When the first boy came along, we named him Allen Ladd—not after the movie star; this is spelled Allen and not Alan—but Ladd was my wife's maiden name. I can remember Allen's brother always thought it should have been Craig and Hogg. The reason that it was Hogg and Craig was that

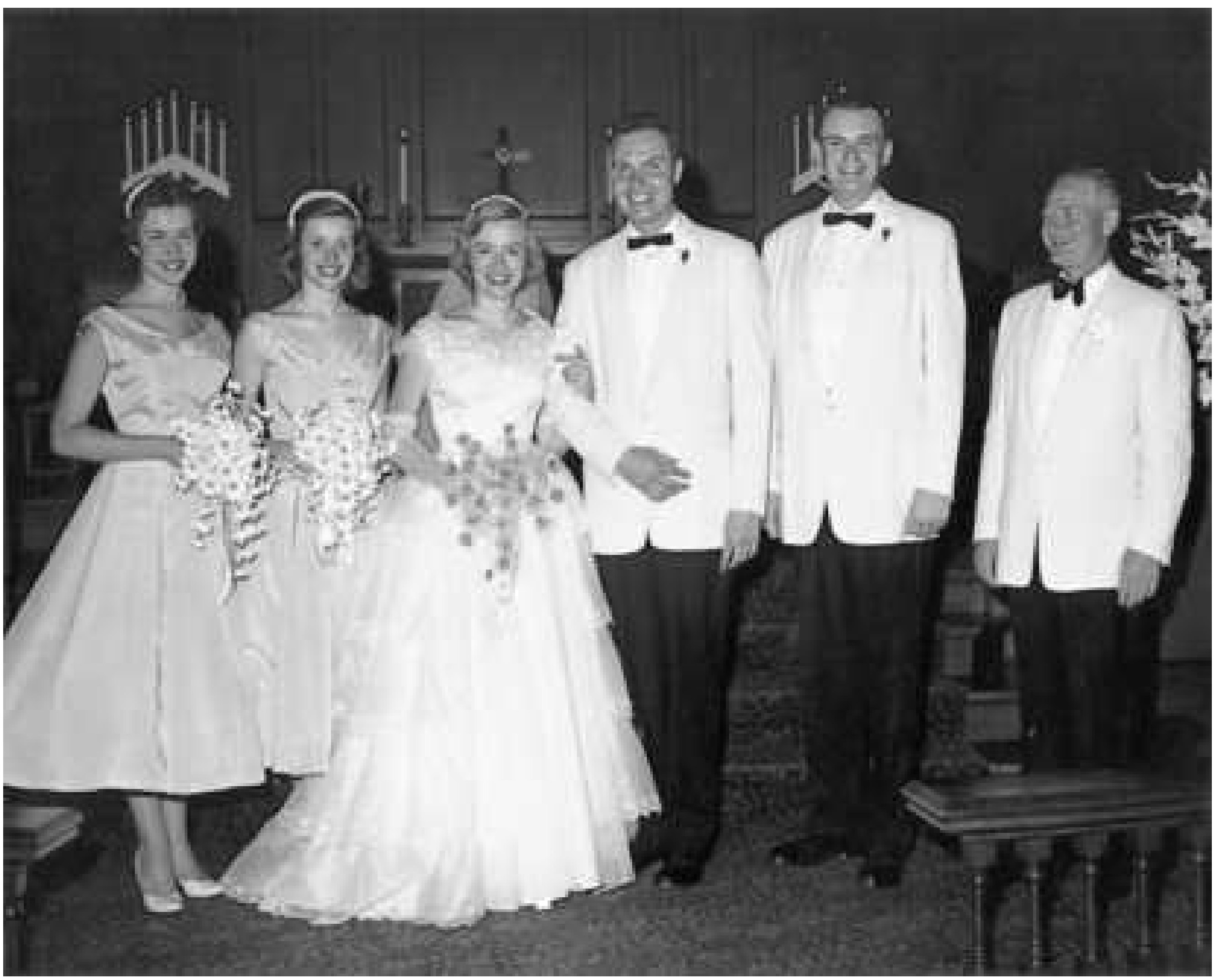

FIG. 3. *Wedding of Carolyn and Bob Hogg, June 23, 1956. Allen Craig is far right.*

Allen was always promoting me. And he said, "Oh Bob, when you revise, I won't help you out." But he helped out on the second one and a little bit on the third edition. But I remember when my son was named Allen Ladd Hogg and his brother said, "Well, Allen finally made first place."

**Randles:** My memory of Allen Craig is that he was a very gentlemanly, kind man. He had the reputation of being a master classroom lecturer. It was described as though his lectures were orchestrated, they were so carefully done.

**Hogg:** There is some truth to that. Somebody used to say that he would come into the classroom and start on the far side of the board away from the door. He would call on students a lot. When he would call on a student, you were it for the day and you just put your pencil down and paid attention. I know he used to pick on me a lot, when he wanted to make time. With some of the students it was like pulling teeth to get the answers out of them. He said, when he would call on me, I would give him the answers quickly. But they'd say he would start on that far board, fill it up, erase it, fill it up again, erase it, and right at the end of the hour he would put the last period in, he would be over by the door and would walk out. His sentences were perfect. I think it was probably due to the fact he was a classics major as well as a math major. There is a story that Allen used to tell about himself. He was in this Greek class and the instructor called on him and he went up and wrote the material out on the board. Allen was a little cocky sometimes. You could tell as he was walking back that he thought he had done extremely well. The Greek professor noticed this attitude and after Allen sat down, the Greek professor just walked to the board where Allen's handwritten work was, put a little accent mark in there that Allen had forgotten to put in. He turned to the class and said "Details mark the master." That is one lesson Allen learned. When we were writing Hogg and Craig or even a paper together, he would come back the next day and he'd say, "Have we considered this or that?" He would be thinking about those details all the time. I had a great relationship with Allen. Carolyn, my wife, was very good to him. She frequently had him over for Sunday dinner or other occasions. He got to know our kids well.

## THE STATISTICS PROGRAM AT THE UNIVERSITY OF IOWA

**Randles:** The Department of Statistics at the University of Iowa was established in 1965 and you were its founding Chair. Tell us about its beginnings.

**Hogg:** I was the one who pushed it starting around 1960. I thought we were falling behind other statistical groups. At that time we had a fairly good relationship with other groups around campus, certainly with the biostatistics group, economics and particularly with educational measurement and statistics. That was primarily due to a fellow by the name of E. F. Lindquist and some of his disciples like Paul Blomers, Leonard Feldt and H. D. Hoover. So for three years starting in 1962, we had an interdepartmental program in statistics so that we could award graduate degrees. We only had one Ph.D. graduate from the program during those years because we were still having Ph.D.'s in mathematics at that time. We added John Birch to the faculty in 1964 and that gave us three statisticians (Craig, Hogg and Birch) and two actuaries (Lloyd Knowler and Jim Hickman). We were able to establish a department with those five people. I remember the Board of Regents, which oversaw Iowa State and Iowa, raised the question of duplication of programs. Iowa State had a great Department of Statistics. I remember that Ted Bancroft, who was the Head of the Iowa State Department at the time, was asked about this by the Regents. His reply was "I can't imagine a major research university without a Department of Statistics." That just shot down the duplication thing. We added Tim Robertson in 1965, Jon Cryer in 1966, Fred Leone was in there, of course you came in 1969. But we went from five faculty in 1965 to around 12 in 1970. As I look back at some of the Ph.D. students we had in the '60s, these were some of the best Ph.D. students we have had: Jim Hickman, Elliot Tanis, John Hewitt, Tom Hettmansperger, Dick Dykstra, Ed Wegman and Doug Wolfe. This was a great collection of students. We were expanding the faculty rapidly. For one thing Sputnik came along and then the NDEA Fellowships, so money was flowing at that time. We had a good department.

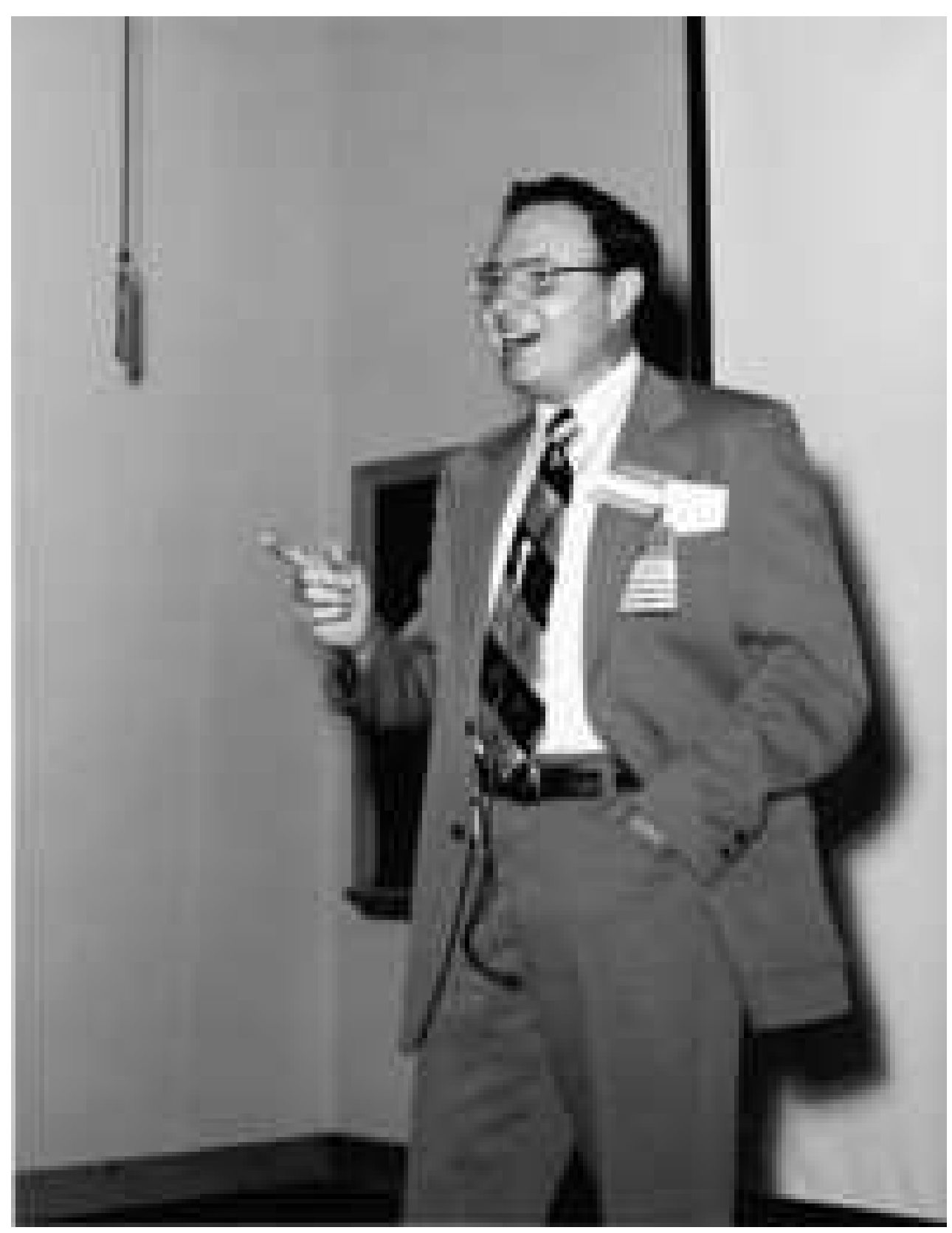

FIG. 4. *As he himself puts it, "Bob Hogg doing his favorite thing—talking" at Los Alamos early in the 1970s*

**Randles:** The Department at Iowa has always had Actuarial Science as part of its mission. Tell us about this association and how it has functioned.

**Hogg:** We started as a Department of Statistics, but later on we changed the name to the Department of Statistics and Actuarial Science. We always had that strong relationship to actuarial science that even went back to Henry Reitz. Henry Reitz in the early part of the last century was at the University of Illinois. He was trained in algebra, but he became a statistician and an actuary. Iowa hired him in 1915 and it was from that day on statistics and actuarial science had always been tied together. Reitz had a lot of Ph.D. students, for example, Sam Wilks, Allen Craig, Frank Weida. One of his actuarial Ph.D.'s was Lloyd Knowler who later joined our faculty. I think we now have 19 faculty members in the Department at Iowa, four of whom are at least an Associate in the Society of Actuaries. We have a strong actuarial program. Most of our undergraduate majors are actuaries. We get a few statisticians, but not many. Our Masters degree in actuarial science is also strong. Occasionally we will get a Ph.D. out of the actuarial group. Elias Shiu, who is our lead actuary, is an actuary specializing in mathematical finance. If he has a Ph.D., the banks or Wall Street want that person. They are in great demand.

**Randles:** You were a member of a select group of professional statisticians in your generation who had the privilege of founding and developing departments of statistics. This was certainly a focal point of your professional life.

**Hogg:** Your comment reminds me of many friends who did help establish departments of statistics: Ralph Bradley at Florida State, Ted Bancroft of Iowa State, Shanti Gupta at Purdue, Frank Graybill at Colorado State, and many more. Possibly some of us stayed too long as executives of these departments, but we were dealing with many young and developing statisticians and it would not have been fair to dump those administrative duties on them. It was a great deal of work dealing with the administrations of the universities. At Iowa I wanted to make certain that statistics and actuarial science maintained their close relationship. It was the right decision to push for a department in the early 1960s because statistics was growing and Iowa would have been left in the dust. I am certain many of my friends throughout the country felt the same way when they helped create statistics departments. The profession is better off because of their efforts.

## INDEPENDENCE

**Randles:** One of the major themes in your research and writing has been independence. You described earlier its development in the Hogg and Craig book [11]. It has been a continual theme really through much of your research.

**Hogg:** Oh, yes. I would joke back in the '50s and '60s that I was "the fastest gun in the west" in recognizing independence or uncorrelated statistics. Now part of that was from Craig who was interested in independence and when he suggested a dissertation topic to me, it was about a ratio,

$$\sum (X_{i+1} - X_i)^2 \Big/ \sum (X_i - \bar{X})^2.$$

When sampling from a normal distribution, that ratio is independent of its denominator. In the early '40s a man named J. D. Williams proved that, but also John von Neumann had a publication in the *Annals* on that. Economists use that result because obviously you can spot a trend because adjacent observations will be close to each other compared to the denominator. That is what started me on my thesis: "On the stochastic independence of a ratio and its denominator." The best thing that I did was a characterization of the gamma distribution, by showing that the independence of

$$\Big(\sum a_i X_i\Big) \Big/ \Big(\sum X_i\Big)$$

and its denominator means you must be sampling from a gamma (or a negative gamma) distribution. I published that in '51 in the *Annals* [6], my first publication. Allen and I didn't do much together my first year on the faculty. He would always go to Florida during the summer time because he had a little ranch down here on which he had some cattle. When he came back to Iowa in the fall of '51, he noted that when you are sampling a normal distribution, if $x$-bar is independent of another statistic, that other statistic has a distribution free of $\mu$. Now this was in the fall of '51. So we started working on that idea. Koopman and Pitman had proved that if you had a single sufficient statistic for a single parameter, you had to have an exponential type distribution. So we started with that and found that the distribution of the sufficient statistic had to be complete; then every statistic which has a distribution free of that parameter is independent of the sufficient statistic. This was a special case of Basu's theorem. In 1952, I submitted a paper concerning the independence of a single sufficient statistic for a single parameter and another statistic. The referee generalized it. Obviously whoever the referee was knew about completeness, and so it wasn't restricted to a single sufficient statistic for a single parameter; you could have a lot of parameters, as long as you had complete sufficient statistics for those parameters. Basu's theorem only takes three, four or five lines to prove it. The referee had done this and of course we started teaching it at Iowa in the fall of 1952.

## ADAPTIVE INFERENCE

**Randles:** I want to ask you about adaptive inference because that's a topic you created and developed extensively, not only in estimation but in testing. That certainly is a topic that has had a major impact on the field. So if you could tell a little bit about how you got into it and what your thoughts were.

**Hogg:** As you point out, there are two aspects of my work in adaptive inference. The estimation part started in the early 1960s [7] when I proved that if the sample arose from a distribution which is symmetric about the parameter $\theta$, an odd location statistic, like $x$-bar, the median or the midrange is uncorrelated with an even location-free statistic, like the variance, the range or the kurtosis. Moreover, the conditional distribution of the odd location statistic, given the value of the even location-free statistic, is symmetric about $\theta$. It occurred to me [9] that you could construct a weighted average of a number of odd location statistics by looking at a measure of tail length, like the kurtosis, to determine the weights. The weighted average would be an unbiased estimator of $\theta$ with hopefully a lot smaller variance than any single one of them, like $x$-bar. For illustration, if the kurtosis was about 3, put lots of weight on $x$-bar. But if it was 6 or more, use more weight on the sample median or a trimmed mean. It worked!

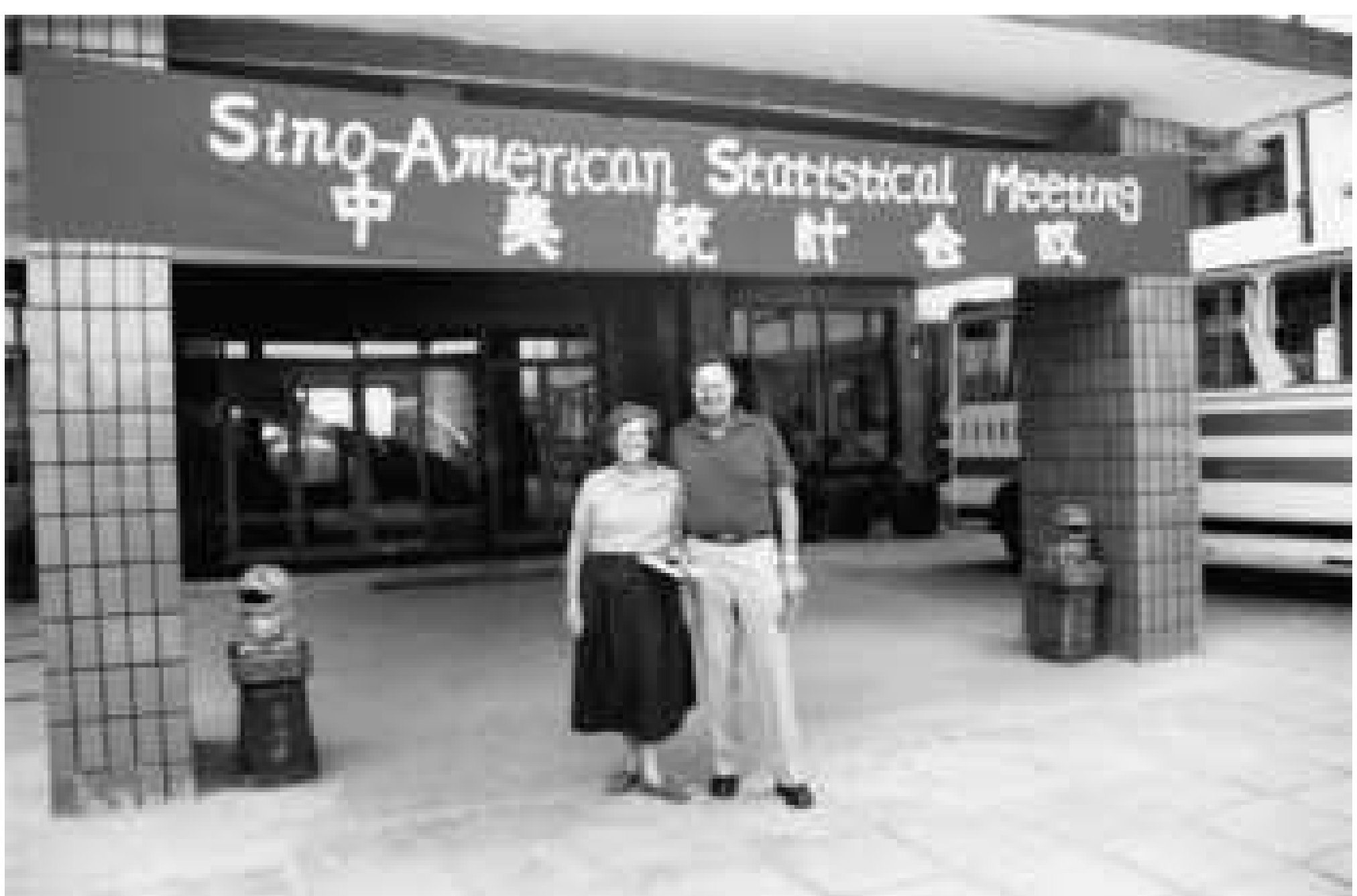

FIG. 5. *Carolyn and Bob at Sino-American Meetings, 1986.*

The testing part started in the early 1960s too [8], although you and I developed it more in the 1970s [12]. Let me explain it using a two-sample test of the null hypothesis $F_1 = F_2$, where these are distribution functions of the continuous type. I think it was in 1963 that I observed that under the null hypothesis, the order statistics of the two samples combined were complete sufficient statistics for $F_1 = F_2$ and thus, by Basu's theorem, were independent of every distribution-free test statistic of $F_1 = F_2$ when the null hypothesis was true. Thus we could use a function of those combined sample order statistics, like the kurtosis, to select which $\alpha$-level distribution-free test to use and still have an overall $\alpha$-level test. This selection allowed us to use a distribution-free test that was more appropriate. Hence for a variety of underlying distributions this adaptive procedure had a much better power than just one distribution-free test statistic, like the Wilcoxon. Of course, you and I know that the Wilcoxon is a very good test for a wide variety of underlying distributions, but it can be beaten in power if the null hypothesis consists of the equality of two distributions that differ enough from the distributions around the normal. For illustration, suppose the two were uniform or Cauchy or highly skewed distributions; then other rank tests will perform much better than the Wilcoxon [10].

## TEACHING AND STATISTICAL EDUCATION

**Randles:** You have been a leader in statistical education and teaching issues within the statistics profession.

**Hogg:** I feel that I have really done something for statistical education. First in mathematical statistics, I think Hogg and Craig did have a lot of influence on statisticians in the '60s and '70s. I was elected Chair of the training section in 1973. The training section was not very strong. I remember going to the business meeting and I think that there were four or five people there, in addition to me. I remember that Brian Joiner and Norm Johnson were there. I thought that the word "training" was terrible—it was sort of like training a dog. I proposed at that meeting that we change the name from "training" to "statistical education." So all five or six voted on it and we thought it was a good idea. So I remember telling Leone, the Executive Director, "Fred we changed the name." Well, he said, "Bob, you got to do something other than just vote on it. You have to rewrite its constitution." So I said, "O.K., Fred, I'll do that." I remember saying "give me a model." So I got something to follow and I rewrote the constitution. I became the Chair again in 1983. The thing that impressed me so much was how the activity had increased between 1973 and '83.

Of course you think of people who contributed to statistical education and I want to mention Fred Mosteller. First, Fred made lots of contributions to statistics, but he gave a talk before the NCTM in 1967 or 1968. This was to teachers of mathematics. Of course he was talking about statistics and what you could do with statistics. Teachers came up to him and asked, "What can we do to get statistics into the high schools?" Fred was very instrumental in doing that, and the ASA and NCTM joint committee on statistical education was started. Out of that "Statistics: A Guide to the Unknown" [19] came out. Of course, Judy Tanur had been the editor of that, but Fred was a leader in that effort. As a matter of fact he came out to Iowa and gave the first Craig lecture in 1971. Fred talked about SAGTU. He is another person who believes in "details mark the master." Fred is the only Craig lecturer that I ever had that wanted to go over to see the auditorium in which he was going to talk beforehand. He wanted two overhead projectors set up. He said, "You go to the back of the room, Bob, and listen to me talk. Can you hear me?" But Fred would flip a transparency on one projector and flip one on the other projector and gave a nice talk. As a matter of fact, I remember driving him to the airport afterward and we got to talking about whether the word statistics—was that a good name for our profession? I remember we got to talking about analytic science and things like that. But you know that once we have all of these Departments of Statistics we are sort of stuck with it even though it might not be the best name.

We have correspondence in the office between Henry Rietz and Egon Pearson along the same line from back in the late '20s or early '30s. Although I did make copies of them, I sent the originals over to Ames because they have the archives of statistics and I thought that was a more appropriate place for them. H. A. David read them and wrote me a letter and said that they really appreciated getting them. There is an interesting side story. You see, Pearson had written long hand and he didn't have copies of the correspondence with Reitz. But Reitz kept carbons of his return letters. So Reitz's file contained both his and Pearson's letters. When Pearson was celebrating his 80th birthday, Mel Novick was going to be there and he had us sign a birthday card. After I signed my name I mentioned that I had this old correspondence. A few months later I got a letter from Egon asking if he could have copies of all that correspondence and so I made copies and sent them to him. I think he was trying to write up his memoirs.

**Randles:** I would like you to talk a bit about your philosophy of teaching because I think as a young faculty member I learned a lot from you about teaching from the discussions that you prompted in the coffee room over lunch. I don't think that young faculty these days get a lot of opportunities to talk about teaching the way we did. I think it was very beneficial.

**Hogg:** I was probably pretty lucky in the fact that I had Craig as a mentor as he was a beautiful lecturer. Allen was always very nice to people and he didn't demean the student. He was very kind and considerate but I think the reason he wanted to call on a student in class was to be sure he was going at the right pace. I tried to teach like Craig and write sentences on the board my first year or two but you have to establish your own style. I did call on students but again in a kind sort of way; I just wanted to see if they could follow through with what they were doing. As a matter of fact, I used to say "Allen, you spoon-feed them too much. You are making it too easy for them. You have to give them a little more to do." I always wanted to give more homework than Allen did. They are going to learn by doing things themselves. I think I was always kind to students and I welcomed students to come in and see me. As a matter of fact, when "I failed in retirement" and I went back to Iowa to teach two courses in the fall of 2003, I had lots of students in to see me. I made it very clear to them the first day that they would be welcomed. I found out that there were about five or six of the actuaries that were going to take the exam in November and I said, "You should never have signed up and spent your $100 because we won't have covered some of the needed material before November." So I met with them on separate occasions. We went to the conference room and I spent an hour and a half or two hours with them three or four times just to get them caught up. So I was working ahead with these students and I've always been very sympathetic to the students. I've enjoyed the students and I've tried to make statistics fun and tried to make it exciting. You know if

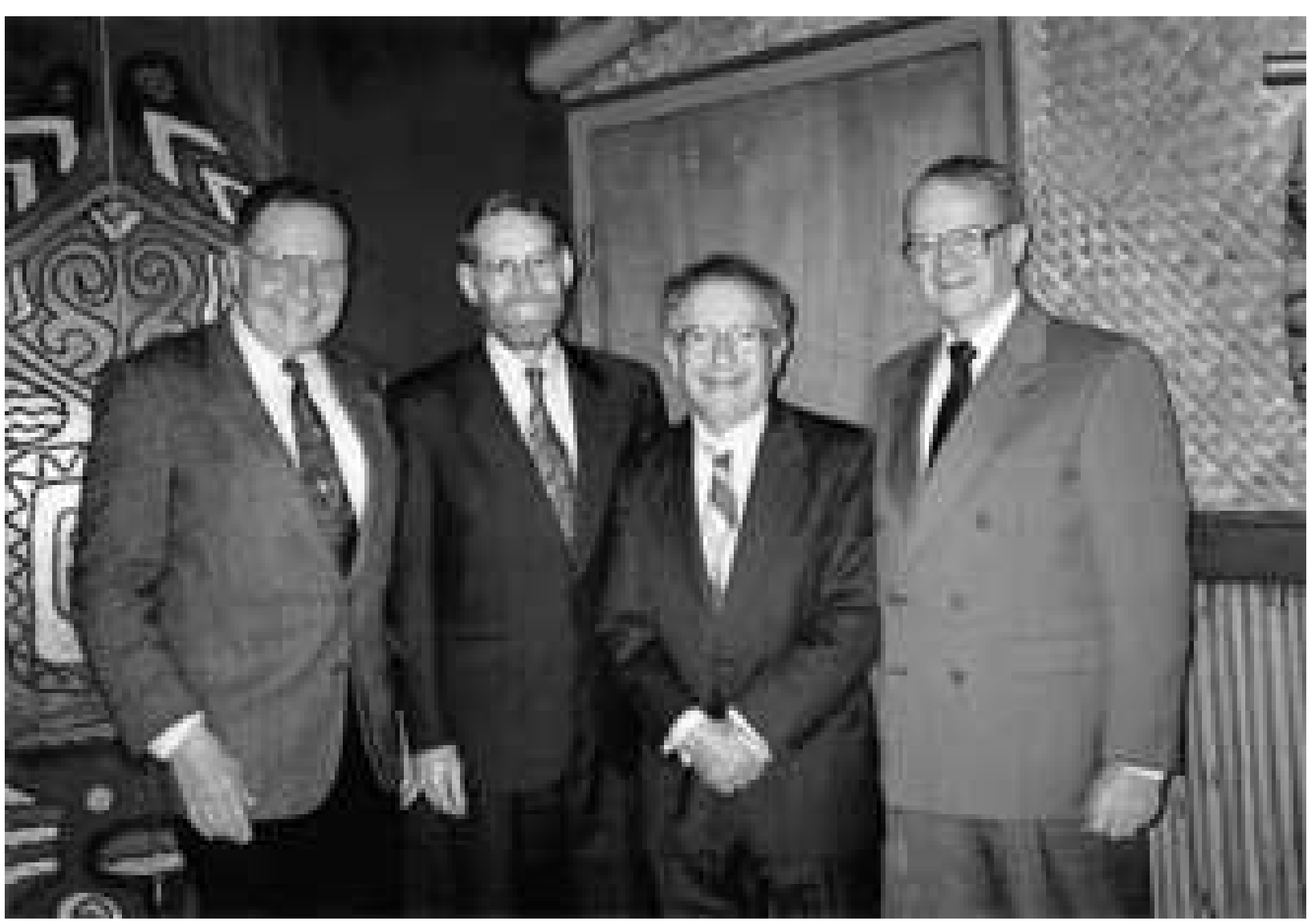

Fig. 6. *Four ASA Presidents: Bob Hogg, Ron Iman, John Neter and Stu Hunter in the early 1990s.*

you are going to lay out all those formulas, it can be pretty dull. You have got to try to give them a little bit of insight. When I went back that fall of 2003, my son said, "Now Dad, be sure you know what stories you have told to which class so you won't repeat yourself." I think people that have taken my courses always thought they were entertaining.

**Randles:** You've received a number of awards for your teaching: the MAA Award for Distinguished Teaching in 1992, the Governor's Science Medal for teaching in 1990 and the Regent's Award for Faculty Excellence in 1992. Obviously you have been recognized for excellence in what you do in the classroom.

**Hogg:** Yes, possibly all those awards started when a young woman who had been in my class in the late '80s wrote an essay about me entitled "There is a hog in my statistics book." It was due to her essay that I received a small local award. But that was the start of it. Then some of my colleagues (Jim Broffitt, Tim Robertson, Dick Dykstra) picked up on that first award and submitted my name for others. It was sort of a snowball effect. I was a good teacher, but probably did not deserve all those awards.

## ASA LEADERSHIP

**Randles:** In addition to your leadership role in statistics at the University of Iowa, you have also been in major leadership positions within our professional societies including the IMS Council, the ASA Board of Directors and serving as President of ASA.

**Hogg:** I did a lot for ASA and I probably deserved the Founders Award. I figured that's OK for what I did. One of the things that I'm very proud of is being President of ASA in 1988. I wasn't a great president. As a matter of fact, I don't remember anything that I really did that was outstanding. I could look at people like David Moore and always think they did more than I did. But, nevertheless, I was President and ASA survived. You know some presidents just disappear and just do nothing after their term is over, but I was so keyed up, I told Lee Decker and Doris Moss that I could run a Winter Meeting for ASA. I was going to have one on statistical education and I did it in January 1992. And we had about 600 people, with about 200 students among those 600. It was in Louisville and it turned out to be a real successful meeting. I had more people say that was the best meeting they ever attended. I had planned the program and arranged for the local people to set up the hotels. Lee and Doris helped out and that was a great meeting. But as far as ASA was concerned they probably thought it was a loser—it was really a winner. Now why was it a loser is because Doris and Lee spent so much time on it and they didn't collect enough at the door to make it compensate for their time. But they were used to running national meetings where they had three or four thousand people and 600 didn't sound like very much. But it was a great meeting and I felt so good about it. They signed my name tag as Boss Hogg from Doris and Lee with appreciation. That one was so good I decided to run the 1994 Winter Meeting on total quality management. Actually that one was in Atlanta and it attracted about 600 people with lots of students. It was a good meeting, but I don't think it was quite as successful as the Stat-Ed one.

**Randles:** You had some tough assignments for ASA. I think, in a way they were hoping you could rescue the Winter Meetings because they were concerned that they had been underattended generally. The one you had in Louisville and the one you had in Atlanta were exceptional in terms of success.

**Hogg:** Yes, those two meetings were good and I am proud that I did not disappear as a past President. Later Ray Waller was the Executive Director and he asked me if I would come in to ASA, free (they paid some expenses) and work on Stat-Ed stuff. This was around 2000. We had a meeting of about 11 or 12 people on Duke Street at ASA; G. Rex Bryce and Dick Scheaffer were there. You know Bryce and Scheaffer really saved me. Those guys could sit there and take notes and type them up on the computer. But actually we had a conference come out of this preliminary meeting, an invited conference. I can't remember where it was held, perhaps Alexandria. Maybe it was a day before a regular meeting that we brought people in.

**Randles:** You have worked with a number of individuals in making important contributions to statistical education through the years.

**Hogg:** Well, of course, I really got interested in statistical education probably through the Hogg and Craig textbook [11] first, but that was more mathematical statistics. Then there were my connections with Fred Mosteller, changing the training section over to statistical education, and then the fact that I got appointed to the ASA/NCTM committee. There was a fellow named Jim Swift who was on the NCTM side and he was energetic. I served on the committee from 1974 to 1977 and I chaired it from 1977 to 1980. Then I went off and Dick Scheaffer was appointed in my place to represent ASA. As I recall, there were maybe four of us from each side, ASA side and NCTM. Dick Scheaffer was a great appointment because Dick, after being on that committee for a few years, started going for NSF grants and he did fantastic in getting money to do all those things. Dick Scheaffer really did a great job for statistical education and I think he is still doing it today. Even though he is retired from Florida, I think Dick is still involved with statistical education. And there's another man that I have to mention and that's David Moore. When David was President of the ASA he made statistical education as one of his main things. I think David decided he was not going to try to do much research, but spend his time on stat-ed. He wrote some good books. David was sort of a godfather of that group of "isolated statisticians." This gang included Tom Moore, Jeff Witmer, Don Bentley, George Cobb and many more. It is a good group that did a lot for "Stat 101" and then advanced placement. But Dick and David were leaders in that amazing growth. Advanced placement started out with about 6000 taking the AP test in statistics and now we are well over 60,000. High school students are learning statistics.

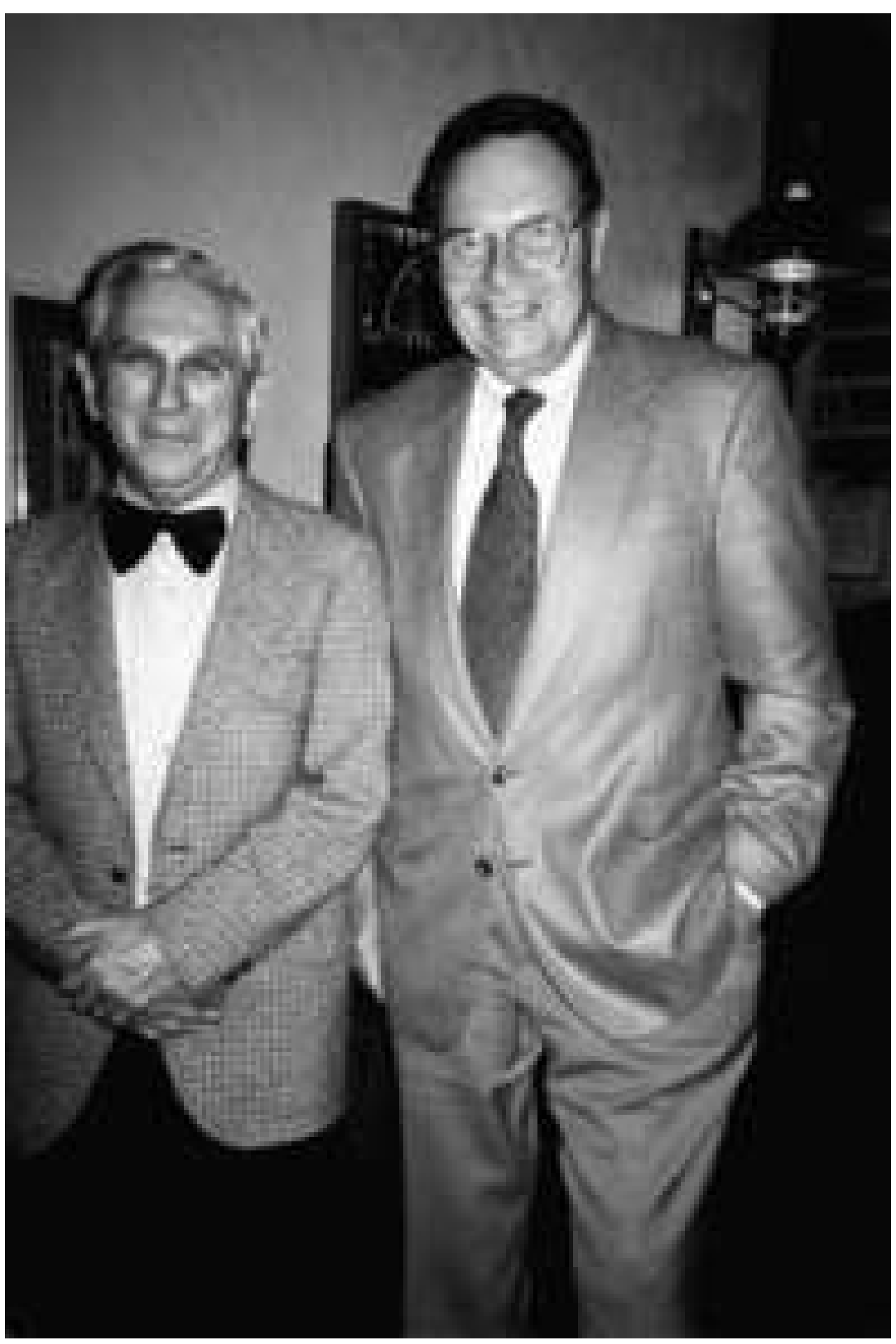

Fig. 7. *Bob with Dick Anderson when Bob gave the first Anderson Lecture at the University of Kentucky in 1996.*

**Randles:** It makes a big difference when they come to the university already knowing about our discipline.

**Hogg:** Oh yes, and they can think about possibly majoring in statistics. Of course in Florida you get quite a few undergraduate majors. In Iowa we get a lot of undergraduate majors although most of them are in actuarial science. But Dick Scheaffer is well known in the overall statistical education effort. Dick is one of those past Presidents of ASA who has kept active in the society. I continue to go to ASA meetings, although I occasionally miss one, like Toronto. I did go to San Francisco, however.

## THE HAM IN HOGG

**Randles:** In a profession of mostly rather timid people, you stand out for your outgoing and gregarious personality. Around the Iowa campus you are known for your antics to "make it fun."

**Hogg:** I have to admit I had fun being a faculty member, and I even went back in the fall of 2003 and did some of my funny things I used to do. I started singing "Thanks for the memories" about 1984. For every class, at the end of the semester I would sing "Thanks for the memories." I would work the students' names into the song, or if I had TA's they would get in, the book we were using got in. Sometimes statistics would get in there, for example the variance and the mean—statistics frequently seen. I remember I sang "Thanks for the memories" in my ASA Presidential Address because I was thanking Fred for being Executive Director. I made just one little verse about it and I sang it during the talk. I'm not a real good singer, but I have guts.

Then, of course, at the end of the fall semester, I would start playing Santa Claus around the campus. I'd wear a Santa Claus suit. As a matter of fact, at first I would rent the Santa Claus suit and then one Christmas my daughter, Barb, made a Santa Claus suit for me. That was the best Christmas present I ever got. So I played Santa Claus around campus. First I would go to class dressed as Santa Claus the last day. Then during final exam week I would go around and see different departments. I'd see the Dean and then I'd go to the President's office. If you are wearing a Santa Claus suit you can get into any meeting you want to. I broke into more darned important meetings. So I did it again in December 2003. We spent some time with my family in the Chicago area and, of course, I dressed up for the grandkids. I played Santa Claus and had all of them sit on my lap. Now they knew it was Grandpa. Some of them were still young enough to believe in Santa Claus and they understood I was just pretending to be Santa Claus. They all sat on my lap and so they got pictures of me.

I would do sort of nutty things like that. I remember being interviewed one time and just out of the blue I said, "My philosophy is anything worth doing is worth doing poorly at first," and the interviewer said, "What do you mean by that?" I explained that I heard so many people say they are going to write a book or they are going to do this or that, and you know they never realized that to write a book you have to pick up a pencil and paper or now go to a computer and do it. So if you have some ideas, write them down. You know you don't have to worry about editing them, but make the first step and write them down. You can edit them later. So that's what I meant by doing it poorly—just get them down. In giving talks, I sometimes say it is better "to vague it up" than to give real proofs. Same philosophy! Anyway, the interviewer quoted me. I have not used that quote since.

## CAROLYN AND ANN

**Randles:** You mentioned earlier that your wife Carolyn was very kind to Allen Craig. I remember her as a very warm and gracious lady who seemed to enjoy the experiences and the travel of your career. Tell us about her.

**Hogg:** Carolyn was an undergraduate English major at Iowa, but she minored in mathematics. That is an unusual combination. She taught school out at Harlan, Iowa. As an English teacher she observed that the math teacher had a lot less correcting to do on tests and things like that. She believed that to teach these kids to write, they had to write essays and to correct those was a terrible job. The math teacher always had it easier. So she decided to go back and get a masters in mathematics. It was about her first year when she was taking a course from me. This was a one-semester course in probability and math stat. She was in my class. Now when she was a student I did not date her, let me make that clear. But when she got her Masters degree, she started teaching math in Cedar Rapids. She was an Iowa City girl and was still living with her parents back in Iowa City. So we started dating. We were married in 1956 and we had four children about three years apart. Allen Craig was the godfather to each of the kids. Carolyn was always really nice to Allen. She would have him over for many Sunday meals. So he was really close to our family. Unfortunately, Carolyn got cancer in the early '80s. She fought it for a long time. She had surgery and went to NIH for special treatment, but she died in 1990. She knew about academics because her father was the Dean of the Law College at Iowa. She was very supportive of my professional work. Those last few years before she died, she was highly dependent on me. I put off several book revisions. After she died, I started working like a young man again. I worked hard on revising the engineering statistics book with Ledolter and on a revision of Hogg and Craig. I worked 60–65 hours a week. But then I started to ease up some; I had had enough of that.

**Randles:** Several years later, you met Ann.

**Hogg:** I met Ann at church. We had coffee together after church. Then finally, when I was 69 years old, we started dating early in 1994. We got married in October of that year. I can always remember telling Ann that I wanted to get married before November 8 because that was going to be my 70th birthday. I wanted to get married while I was still 69. It was a good thing for me. I had been alone there for four years. Ann has been very good for me and she has been very supportive of my professional activities. Whenever I needed to work on things, she was there to support me.

## INFLUENTIAL PERSONALITIES WITHIN THE PROFESSION

**Randles:** Maybe you could tell us about some of the statisticians in the profession who influenced your career or with whom you had memorable relationships.

**Hogg:** You know one person that I never met was Harald Cramér. But I felt that his book had a great effect on me. Now it is out of date as far as statistics is concerned but you've got to remember that it was published in '46 by the Princeton Press. And so at that time that book had a big influence on me. The mathematics in it is great. I have to say when Craig and I were writing our book we would always say, "Let's see how Cramér did it." I'm not saying we plagiarized but let's just see how he did it. That was one influence. I would also like to mention Frank Graybill. Again, of course, I've known Frank for a long time. As a matter of fact, I think the first time I met him was at a Cambridge meeting in the late '50s. When I saw that this was Frank Graybill, I introduced myself. I was going to tell him about writing Hogg and Craig and before I got that out he says, "You know I'm revising Mood." I thought, I'm not going to sell him any Hogg and Craigs. Well, of course, afterward I had become pretty good friends with Frank and I kidded him about it and said, "Now, Frank, you messed up Mood enough so that you gave Hogg and Craig a chance." He always laughed.

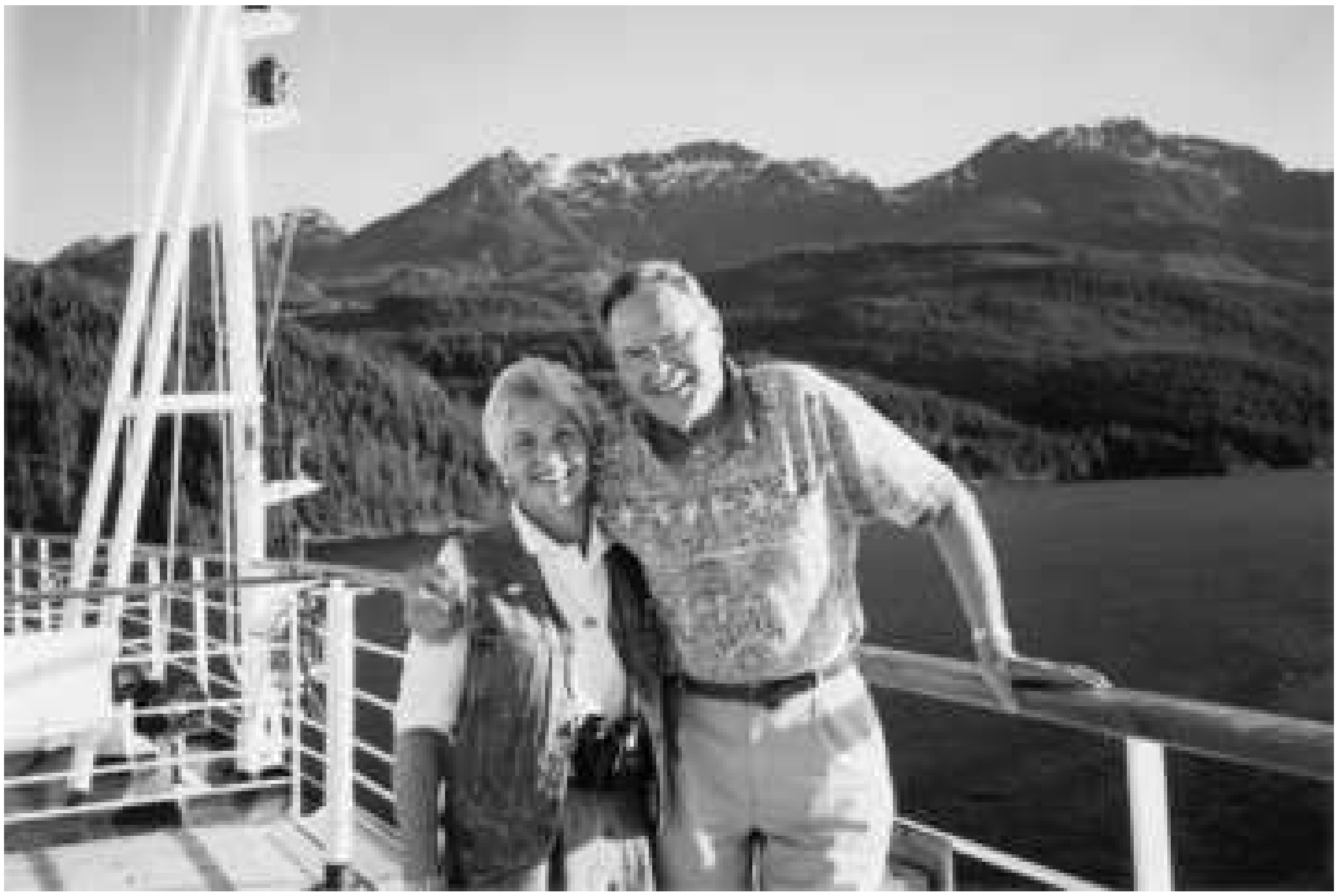

Fig. 8. *Bob and Ann on a cruise in the late 1990s.*

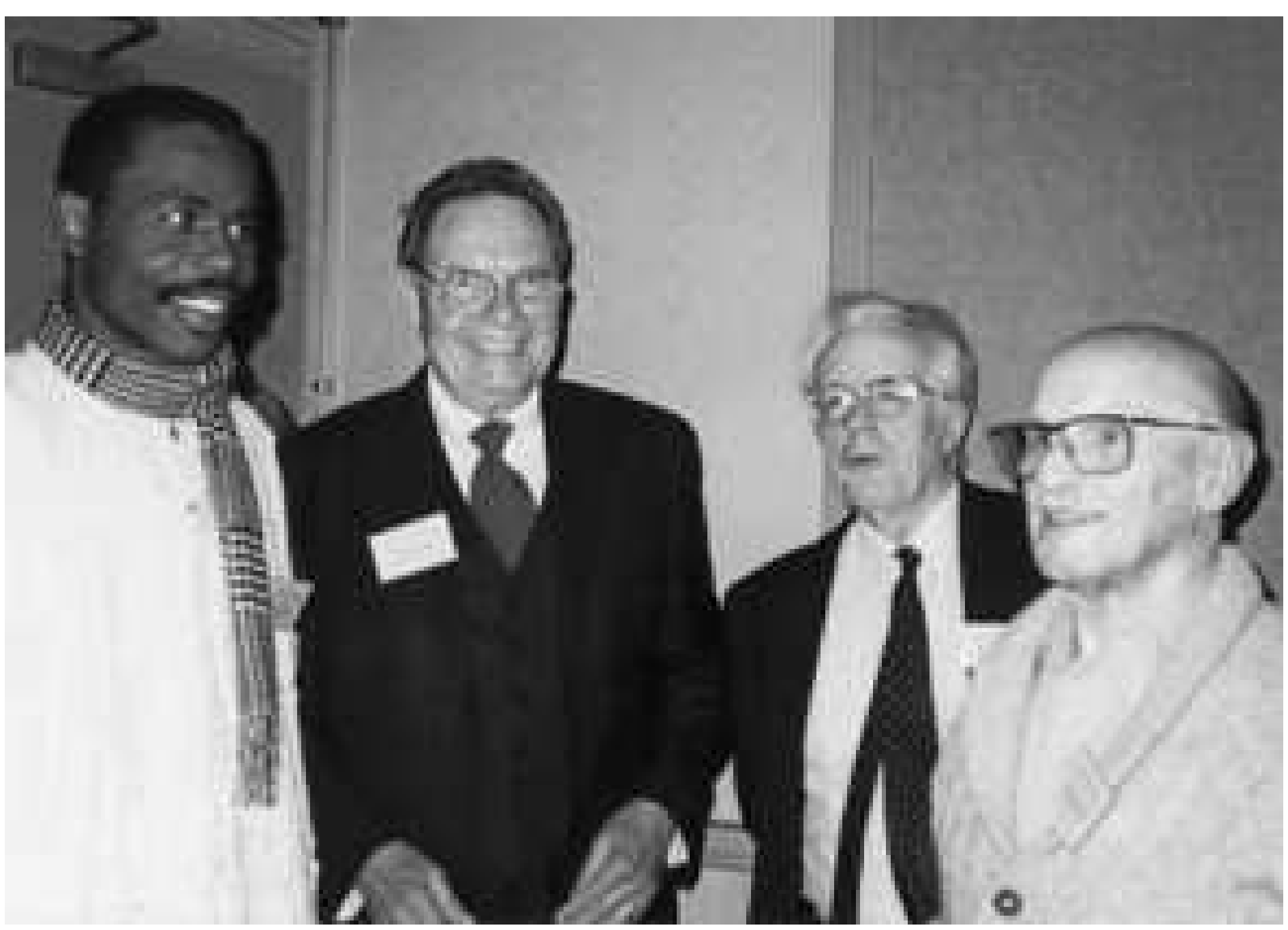

Fig. 9. *A Graduate Student, Bob, Sir David Cox and Oscar Kempthorne at H. A. David's retirement dinner in 1995.*

Oscar Kempthorne and I had a great relationship and I remember him coming to my 60th birthday party. Oscar got up and said a few things and I went up to thank him and I kissed him on the top of his head. I thought it was kind of a neat thing, but I think some Iowa State'ers thought I shouldn't have done that. Iowa always had a great relationship with the statisticians at Iowa State. Ted Bancroft was the first Head and I didn't know much about Iowa State in the early '50s but Ted and I got pretty well acquainted in the late '50s and '60s. We started a joint seminar between the two departments. Ted always tried to make it a point to be there. He was highly supportive of that joint relationship. Of course, after Ted they hired H. A. David to be Head at Iowa State. H. A. and I hit it off. His wife, Vera, died about the time that my wife, Carolyn, had died, so when I would see him I always said, "How you doing, Herb?" Sure enough, one meeting I went up to him and I said, "How you doing?" He said, "I got remarried". This was to Ruth, who was a friend of Vera's, as I recall. Ruth is a lovely lady, as was Vera. I figured if H. A. could get remarried, so could I, and Ann and I were married about a year later in 1994.

I'm going to mention something about George Box because he's a neat guy and very interesting and I love that guy. One time I was with George and this was after I had written the engineering statistics book with Hannes Ledolter [14]. I said, "George, we just copied some of your designs." He said, "Well, you could have done worse" or something like that. Of course, the reason we copied them was the fact that Ledolter was a student of George's. George is a neat guy and a very interesting man. I think Box and Stu Hunter are in the process of revising Box, Hunter and Hunter.

Of course, Stu is another good friend of mine. I always remind Stu, who was the first editor of *Technometrics*, that I had sent a paper into the journal and that he had immediately rejected it. Now the reason he rejected it is that you had to have data in articles in that journal. I didn't have any data in there. Stu just rejected my paper right off the bat. Later on I got to know Stu very well. As a matter of fact, when I was President of ASA, I was trying at that time to get one of the quality gurus to give the Presidential Invited Address. I talked to Stu and I told him I was having a problem getting a speaker. I only had one more man to try. But if I get a no, I'm going to ask you to give my invited address and that's the way it turned out. So Stu is a great friend. When I think of Stu, I remember when he got Deming for his invited talk. This was the '93 San Francisco meeting. That was in August and Deming died in December of that year. I remember Deming talked on systems but he gets rambling, more than I do. He rambled about all these things and it was more of an event than a talk. I can remember afterwards some guy asked him, "What role is the computer going to play in the quality movement?" Deming answered, "This is a serious discussion, next question." Then there was a man who met him at the end of the talk who had written a book and wanted Deming to have a copy of it. Deming said, "No, no, no." Finally Deming, to get rid of him, took the book. He was in a wheelchair. Stu had wheeled him on the platform and was going to wheel him off. The book was on the floor. Stu said, "Oh, Dr. Deming, you lost your book, it's on the floor." "Leave that book on the floor," said Deming. Good old Deming, that was typical Deming.

I also want to mention Fred Leone. He joined our faculty before you did in about 1967. He was a good applied statistician and actually I had kind of hoped he would get into the quality control thing a little more than he did, but Lloyd Knowler was very protective of SQC. You know Lloyd took John Ramberg under his wing. He never asked Fred to do any of the SQC activities. Fred did a good job of teaching and he helped us get a nice NSF grant. Actually I nominated Fred for Executive Director of ASA. Fred wanted it and eventually got it. Fred was always a good friend, and I stayed in his home in Silver Spring when my wife, Carolyn, was at NIH having her cancer treated in the late 1980s. Betty, Fred's wife, died a few years ago. Fred was Executive Director from the early 1970s to the late 1980s and I would say he took ASA from a mediocre society to being a world-class organization.

## HONORS

**Randles:** Earlier in this conversation, you mentioned that you had received the Founders Award from ASA. I know that was a great honor. But more recently you received the Gottfried Noether Award from ASA for contributions to nonparametric statistics and you are also slated to receive the Carver Medal from the IMS.

**Hogg:** When I was given the Noether Award, it was one of the highlights of my career. Lehmann had received it and obviously he was everybody's choice for the first one. Well, this was the second time the award was given and I was honored because I thought I knew Gottfried really well. I was very pleased but as I told you there were other people, probably Pranab Sen should have gotten it before I did; also Myles Hollander should have gotten it. I really should have been far down on the list because Pranab, Myles, Tom Hettmansperger, Jay Conover and others are really better than I am. You know I was very pleased. I have to say that I never thought that I was quite deserving of it. I'm glad Pranab was the next and Myles Hollander was the next year and then Tom Hettmansperger. I really was honored because I felt so close to Gottfried and his wife.

I had gotten to know Gottfried when Fred Leone appointed me to the ASA/ NCTM committee in 1974 and Gottfried Noether was also on the committee. Gottfried and I never talked about nonparametrics except Gottfried was strong on the idea that you ought to use ranking to teach elementary statistics. He said it was easier and you got good efficiency. We talked about teaching statistics that way. But we never talked about a research problem in nonparametrics; it was always statistical education. We got to be very good friends.

The Carver Medal came out of the blue just recently. They established this award about six or seven years ago to recognize persons who had made important contributions to the IMS. I had done a lot of work for the IMS in the past as Associate Program Chair and then Program Chair from 1968 to 1974. I also served on the IMS Council as Program Chair. I had just done this work for the good of the profession. Some people recognized that effort and decided to give me this special honor. I am certainly grateful. Establishing an award named for Carver was a good idea because I always thought Harry Carver got cheated. Henry Reitz was recognized. Of course Reitz should have, he was the first President of the IMS and really started the IMS in 1935. But Carver started *The Annals of Mathematical Statistics* in 1930 at his own personal expense. He got a little help from ASA at first. But during the depression, about 1932 or '33, ASA decided to drop its support. Carver kept the journal going.

## CLOSING REMARKS

**Randles:** Do you have any comments that you would like to make as we conclude our conversation?

**Hogg:** I look back and I think I never could have done it alone. I think first of Carolyn and Ann and then some of my professional colleagues. Of course, I have to put Allen Craig at the top of my list, but, Ron, you and I had a good research partnership. Then there are Elliot Tanis [16], Hannes Ledolter, Joe McKean [15] and Stuart Klugman. I wrote a book on loss distributions with Stuart [13] primarily for the actuaries and the casualty people in particular. It dealt with long-tailed distributions. I think of all my colleagues; Tim Robertson was a great help, Jon Cryer and many others who helped me out. That makes me think about my little friend Statistics. Allen and I would go for coffee sometimes in the late afternoons and we would always bring Statistics along. She was a cheap date and we would talk about her. I used to tell people that initially I was shy when I went to professional meetings. Allen didn't go to many professional meetings after World War II. So I didn't have anybody to introduce me. The other day I listed a whole bunch of people who I had the privilege of knowing, for example, C. R. Rao, Sir David Cox, Jerzy Neyman, George Box, Ingram Olkin, David Blackwell, Gottfried Noether, Stu Hunter, Jack Keifer and on and on. I was on a first-name basis with these giants of the profession. I could keep listing people and wonder how did I get to meet those people? Well, my little friend Statistics introduced me. I have been lucky that way. As I look back, I just hope other statisticians can make such a friend of Statistics and have the relationship that I have had with Statistics. It was a good life.

## ACKNOWLEDGMENTS

We express our thanks to Allan Sampson for suggesting this interview and to the Department of Statistics at the University of Florida, and especially, the Chair, George Casella, for facilitating the opportunity. We also express our thanks to the Executive Editor, Ed George, and an Editor for their constructive suggestions.